\documentclass[sigconf, table]{acmart}

\usepackage{booktabs} 
\usepackage{tabulary}
\usepackage{rotating}
\usepackage{enumitem}

\copyrightyear{2021}
\acmYear{2021}
\setcopyright{acmcopyright}\acmConference[GECCO '21]{2021 Genetic and Evolutionary Computation Conference}{July 10--14, 2021}{Lille, France}
\acmBooktitle{2021 Genetic and Evolutionary Computation Conference (GECCO '21), July 10--14, 2021, Lille, France}
\acmPrice{15.00}
\acmDOI{10.1145/3449639.3459285}
\acmISBN{978-1-4503-8350-9/21/07}

\begin{document}

\title{PSB2: The Second Program Synthesis Benchmark Suite}

\author{Thomas Helmuth}
\orcid{0000-0002-2330-6809}
\affiliation{%
  \institution{Hamilton College}
  \city{Clinton} 
  \state{New York} 
  \country{USA}
}
\email{thelmuth@hamilton.edu}

\author{Peter Kelly}
\affiliation{%
  \institution{Hamilton College}
  \city{Clinton} 
  \state{New York} 
  \country{USA}
}
\email{pxkelly@hamilton.edu}



\begin{abstract}
For the past six years, researchers in genetic programming and other program synthesis disciplines have used the General Program Synthesis Benchmark Suite to benchmark many aspects of automatic program synthesis systems. These problems have been used to make notable progress toward the goal of general program synthesis: automatically creating the types of software that human programmers code. Many of the systems that have attempted the problems in the original benchmark suite have used it to demonstrate performance improvements granted through new techniques. Over time, the suite has gradually become outdated, hindering the accurate measurement of further improvements. The field needs a new set of more difficult benchmark problems to move beyond what was previously possible.

In this paper, we describe the 25 new general program synthesis benchmark problems that make up PSB2, a new benchmark suite. These problems are curated from a variety of sources, including programming katas and college courses. We selected these problems to be more difficult than those in the original suite, and give results using PushGP showing this increase in difficulty. These new problems give plenty of room for improvement, pointing the way for the next six or more years of general program synthesis research.
\end{abstract}



\maketitle

\section{Introduction}








Automatic general program synthesis, with the aim of automatically generating programs of the type humans write from scratch, has long been a goal of artificial intelligence and machine learning. Yet, for many years there were no common benchmark problems for evaluating general program synthesis\footnote{Also known as automatic programming or software synthesis.} systems; existing problems were either easy toy problems or were situated in specific domains where solution programs were composed of a small set of domain-specific instructions.
In 2015, the General Program Synthesis Benchmark Suite (PSB1) introduced 29 problems that could be used to benchmark program synthesis systems~\cite{Helmuth:2015:GECCO}. Since then, more than 80 research papers have benchmarked 10+ program synthesis systems using PSB1, producing numerous insights into program synthesis.

Of the systems that have adopted PSB1, most fall within the field of genetic programming (GP), including PushGP~\cite{Helmuth:2015:GECCO}, grammar-guided GP~\cite{Forstenlechner:2017:EuroGP}, grammatical evolution~\cite{Hemberg:2019:GECCO}, and linear GP~\cite{Lalejini:2019:GECCOcomp}. However, non-evolutionary program synthesis methods have also been applied to PSB1, including those based on delayed-acceptance hillclimbing~\cite{DBLP:journals/corr/abs-1811-10665} and Monte Carlo tree search~\cite{lim2016field}. We expand on the details of these methods and the results they have achieved using PSB1 in Section~\ref{section:past-research}, but to summarize, many of these systems have improved performance and demonstrated new techniques.

When PSB1 was first introduced, the initial PushGP runs were able to solve 22 of the 29 problems, with an average success rate of 23 successful runs out of 100~\cite{Helmuth:2015:GECCO}. The best-performing PushGP results have now solved 25 problems, with an average success rate of $42/100$~\cite{Helmuth:2020:ALife:downsampledlexicase}. Some of the most drastic improvements have come on some of the most informative problems in PSB1, such as Double Letters (6~$\rightarrow$~50 successes between \cite{Helmuth:2015:GECCO} and \cite{Helmuth:2020:ALife:downsampledlexicase}), Replace Space with Newline (51~$\rightarrow$~100), Syllables (18~$\rightarrow$~64), Vector Average (16~$\rightarrow$~97), and X-Word Lines (8~$\rightarrow$~91).

Thus, for PushGP and other synthesis systems, the problems of PSB1 have become less useful over time. In particular, the very high performance achieved on some PSB1 problems leaves little room for exhibiting improvement; a few other problems have never been solved and are likely too difficult to be solved any time soon. Additionally, peculiarities in some of the problems in PSB1 make them less ideal as benchmarks, either because of how synthesis systems move through their search space or how slow they are to run. Finally, some decisions about the specification of problems in PSB1 make them difficult to implement in hindsight, potentially preventing wider adoption.

With these drawbacks in mind, we have created a second Program Synthesis Benchmark suite, which we refer to as PSB2. PSB2 consists of 25 problems curated from programming challenges, programming katas, and college courses. In order to facilitate the uptake of PSB2, we provide a reference implementation of each problem, as well as datasets that can be sampled to more easily implement each problem in new synthesis systems. \footnote{Reference implementation, datasets, and other resources can be found on this paper's companion website: \url{https://cs.hamilton.edu/~thelmuth/PSB2/PSB2.html}.}
Just like PSB1, the problems in PSB2 require a wide range of programming techniques, data types, and control flow structures to solve. However, they are markedly harder to solve than problems in PSB1, with our initial results solving 13 of the 25 problems for an average success rate of $10/100$. These more difficult problems will drive program synthesis research toward solving more realistic program synthesis tasks.

The purpose of benchmark problems is to allow us to empirically show what changes to a system produce improvements that may transfer to real-world problems. To achieve this goal, they must be sufficiently difficult, unlike toy problems that have been used as benchmarks in the past. They must also be representative of the types of tasks we want our system to perform. However, we also want benchmarks to be easier and faster to run than an actual real-world problem in order to aid reasonable testing of a system. Given that automatic program synthesis is still in its fledgling stages, we see the problems in PSB2 as a stepping stone toward solving more realistic problems.

PSB2 also addresses calls from the GP community to produce and adopt realistic benchmarks. GP community discussions calling for better benchmarks~\cite{McDermott:2012:GECCO, White:2013:BGB:2441218.2441242, Woodward:2014:GECCOcomp} inspired the creation of PSB1; these calls also highlighted the need to periodically update and replace benchmark problems in order to keep advancing the field without over-optimizing to a single set of problems. More recently, a call to refocus the efforts of GP on automatic programming stated, ``We are in no doubt of the need for the further principled development of additional benchmarks that can be used in a targeted manner to push the boundaries along different dimensions such as scalability, generalisation, and adaptation, and to facilitate comparison across a range of very different approaches to automatic programming''~\cite{ONeill:GPEM20}. The creation of PSB2 aims to push the boundaries of program synthesis research and give synthesis systems a fresh set of problems to explore. Of course, there is no need to entirely throw out the problems of PSB1; we could imagine some of the harder problems continuing to provide useful data, and newer systems may need to start on the easier problems as a jumping off point.

The remainder of this paper is structured as follows: in the next section, we discuss research that has used PSB1. In Section~\ref{section:lessons}, we highlight lessons learned about program synthesis benchmarking from PSB1. Sections~\ref{sec:sources} and \ref{section:problem-descriptions} describe the sources of PSB2's problems and describe the problems in detail. We then give general guidance on benchmarking with PSB2, and give details of the parameters we used in our experiments in Sections~\ref{section:using-psb2} and \ref{section:methods}. Finally, Section~\ref{sec:results} presents initial results using PushGP.

\section{Past Research Using PSB1}
\label{section:past-research}





PSB1 has been used in a variety of research projects on automatic program synthesis, many of them using GP as the synthesis system. The paper that introduced PSB1~\cite{Helmuth:2015:GECCO} used PushGP, a GP system based on the stack-based Push programming language; a variety of papers using PushGP have made use of PSB1 since~\cite{Helmuth:2016:GPTP, Helmuth:2017:GECCO, Helmuth:2018:GECCO, Helmuth:2020:ALife:downsampledlexicase, Helmuth:2020:ALife:source, Saini:2019:GPTP, Saini:2020:GECCOcomp}. Code-building GP is a stack-based GP system borrowing some inspirations from Push that constructs programs in a host language; it solved some of the PSB1 problems, producing solution programs in Python~\cite{Pantridge:2020:GECCO}.

General program synthesis requires the manipulation of multiple data types; stack-based GP systems have handled this requirement well, but so have other GP systems that handle strong typing of programs. In particular, grammar-based approaches such as grammar guided GP (G3P) \cite{Forstenlechner:2017:EuroGP, Forstenlechner:2018:CEC, Forstenlechner:2018:PPSN, Forstenlechner:2018:GECCO} and grammatical evolution (GE) \cite{Hemberg:2019:GECCO, Kelly:2019:EuroGP, ONeill:2019:CEC, Sobania:2020:EuroGP} have made good progress at solving the problems in PSB1. Many of these use the type-based grammar design patterns introduced to flexibly handle problems with different type requirements~\cite{Forstenlechner:2017:EuroGP}. Another use of these grammars trains a sequence-to-sequence variational autoencoder to embed programs in a continuous space and then uses an evolutionary algorithm to optimize programs in this space \cite{Lynch:2020:PPSN}.
Finally, a linear GP system with tag-based memory has also been explored using PSB1 \cite{Lalejini:2019:GECCOcomp, Hernandez:2019:GECCOcomp, Ofria:2019:GPTP}. 

As for non-GP systems, an approach using delayed-acceptance hillclimbing for inductive synthesis proved competitive with GP on PSB1, including producing the only known solutions to the Collatz Numbers problem
\cite{DBLP:journals/corr/abs-1811-10665}.
A comparison was made between Flash Fill~\cite{Gulwani2011}, MagicHaskeller~\cite{Katayama2010}, PushGP, and G3P, finding that the non-GP methods fared much worse but ran much faster than the GP methods~\cite{Pantridge:2017:GECCOa}. Finally, Monte Carlo tree search was used to generate Java bytecode programs using a few of the problems in PSB1~\cite{lim2016field}.

\section{Lessons Learned}
\label{section:lessons}

While PSB1 has been successfully used in a variety of research, it was a first attempt at a general-purpose program synthesis benchmark suite. The research community has grown from using it, both in terms of improving program synthesis methods as well as lessons learned about how to best define program synthesis benchmarks. Here, we discuss some issues of the latter type and how they have influenced our creation of PSB2.

One major issue with PSB1 is that every system that uses it needs to implement all of the problems from scratch. This hurdle likely decreased wider adoption. Additionally, there may be inconsistencies between implementations in different systems, leading to less comparable results; one known such inconsistency is that some systems use new randomized data for each run, while others use the same dataset for every run. Four years after its initial release, the authors of PSB1 created large datasets of the inputs and correct outputs for each problem~\cite{helmuth:github:2019}. These datasets can be sampled for each program synthesis run, meaning there is no need for each system to implement each problem. We have copied this model, and provide datasets for each problem in PSB2 (see Section~\ref{section:using-psb2}).

A handful of the problems in PSB1 require programs to produce Boolean outputs, as such functions are common in programming exercises. A trend noted across completely different program representations is that solutions to these Boolean-output problems often do not generalize to unseen data. A simple explanation for this phenomenon is that it is relatively easy for a solution program to produce the correct answers for the wrong reasons when there are only two possible answers, thus overfitting to the training data. It is much harder to perfectly answer training data for the wrong reasons when the output is an integer or string, for example. Because of this issue, we have selected fewer Boolean-output problems for PSB2, including only one representative problem.

PSB1 was designed to emulate the textbook problems it was curated from as closely as possible. For example, many problems from the original textbook required the program to print its answers. PSB1 suggested that synthesis systems develop methods for emulating an output buffer and common printing instructions in order to mimic these problems. However, this approach was infeasible for some synthesis systems, which instead simply returned string outputs. As PSB2 is less loosely coupled with its problem sources, we decided to have all programs return their outputs instead of ``printing'' them. Another wrinkle related to outputs is that some problems require a solution to return multiple outputs. While multiple outputs may prove difficult in some systems, it is generally feasible in all; we have included 4 multi-output problems in PSB2.

PSB1 recommended different training and test set sizes, as well as program evaluation budgets, for each problem. This led to difficulties and confusion in both implementation and reporting results. For PSB2, we recommend using a fixed setting for each parameter across problems, as well as for system-specific parameters such as maximum sizes for program size control.

Forstenlechner et al. \cite{Forstenlechner:2018:CEC} discussed understanding and refining the problems in PSB1, making some general recommendations about both synthesis systems and benchmark problems. One suggestion put forth is using larger and more targeted training sets, to better guide synthesis and increase generalization. We take this recommendation and use large training sets (200 examples) that have a variety of specific edge cases purposefully included. Most of their other suggestions relate to specific system settings, such as the length of evolution; these parameters are not prescribed by PSB2, and can be chosen by the researcher.

\section{Problem Selection and Sources}
\label{sec:sources}

Below we describe the four sources we used as inspiration for the problems included in this suite. Each of these sources presents problems for humans to use to improve their programming skills, whether for experienced programmers or students in class. As such, these sources contain problems representative of the types of programming that we expect humans to perform.

\textbf{Code Wars (CW)} - A website full of user-created programming challenges, called \textit{coding kata}. The aim of the site is for users to spend small amounts of time programming every day to hone their coding skills.

\textbf{Advent of Code (AoC)} - An Advent calendar of coding problems created every year in December. These problems can be used for any number of things, like training, interview prep, or coursework. Problems tend to become harder throughout the month.

\textbf{Homework Problems (HW)} - These problems come from programming homework given in our undergraduate programming courses. The problems come from two courses: an introductory programming course, and a program languages course. These problems do not have citations, since we created them for our courses.

\textbf{Project Euler (PE)} - A website containing hundreds of problems in an archive. Users are free to submit answers to validate their solutions. Most problems tend to be mathematically focused and often require efficient and elegant solutions.

These sources contain a large number and variety of problems; we considered over 75 problems from these sources and implemented and tested over 50 of them, filtering out problems that seemed too easy, too difficult, or inaccessible. While curating the suite, we did not include any problems on which PushGP produced a success rate over 60\%, to ensure that problems are sufficiently difficult to allow for improvement. In order to be transparent about our curation process, we have created a table containing all of the problems we considered, including the reason for rejection for rejected problems, initial results if we implemented the problem, and a link to the source of the problem.\footnote{\url{https://docs.google.com/spreadsheets/d/e/2PACX-1vQKO1D2sZA9KosXpOJNuiDW6yDQEZnMrwzNeMJJU25MbZhU6odQ0jGkJN5lgbRgspsmmum65WLbEI2B/pubhtml?gid=0&single=true}}

We aimed to include problems that require a large variety of data types and control flow structures to solve, with a balance between data types across problems. Most, if not all, of the problems require some type of iteration and/or conditional execution. Required data types include integers, floats, Booleans, characters, strings, vectors of integers, and vectors of floats. In order to produce large datasets, we aimed to select problems that have at least 1 million possible unique inputs; the only exception being the Coin Sums problem, which has 10,000 possible integer inputs.

\section{Problem Descriptions}
\label{section:problem-descriptions}

Below is a list of the English language descriptions of the 25 benchmark problems in PSB2. Each problem (besides those from our courses) has a citation of its source with a link to the original problem. The types of the input(s) and output(s) for each problem are given in Table~\ref{table:inputs-outputs}. For more precise details of each problem, see the reference implementation.\footnote{\url{https://github.com/thelmuth/Clojush/releases/tag/psb2-v1.0}}

\begin{table*}[t]
\centering
\caption{For each problem, the types of the inputs and outputs, and the limits imposed on the inputs.}
\label{table:inputs-outputs}
\rowcolors{2}{gray!15}{white}
\begin{tabular}{l p{11cm} l}
    \toprule
    \textbf{Name} & \textbf{Inputs} & \textbf{Outputs} \\
    \midrule
    Basement & vector of integers of length $[1, 20]$ with each integer in $[-100, 100]$ & integer \\
    Bouncing Balls & float in $[1.0, 100.0]$, float in $[1.0, 100.0]$, integer in $[1, 20]$ & float \\
    Bowling & string in form of completed bowling card, with one character per roll & integer \\
    Camel Case & string of length $[1, 20]$ & string \\
    Coin Sums & integer in $[1, 10000]$ & 4 integers \\
    Cut Vector & vector of integers of length $[1, 20]$ with each integer in $[1, 10000]$ & 2 vectors of integers \\
    Dice Game & 2 integers in $[1, 1000]$ & float \\
    Find Pair & vector of integers of length $[2, 20]$ with each integer in $[-10000, 10000]$, integer in $[-20000, 20000]$ & 2 integers \\
    Fizz Buzz & integer in $[1, 1000000]$ & string \\
    Fuel Cost & vector of integers of length $[1, 20]$ with each integer in $[6, 100000]$ & integer \\
    GCD & 2 integers in $[1, 1000000]$ & integer \\
    Indices of Substring & 2 strings of length $[1, 20]$ & vector of integers \\
    Leaders & vector of integers of length $[0, 20]$ with each integer in $[0, 1000]$ & vector of integers \\
    Luhn & vector of integers of length 16 with each integer in $[1, 9]$ & integer \\
    Mastermind & 2 strings of length 4 made of \texttt{B, R, W, Y, O, G} & 2 integers \\
    Middle Character & string of length $[1, 100]$ & string \\
    Paired Digits & string of digits of length $[2, 20]$ & integer \\
    Shopping List & vector of floats of length $[1, 20]$ with each float in $[0.0, 50.0]$, vector of floats of length $[1, 20]$ with each float in $[0.0, 100.0]$. Both vectors must be the same length & float \\
    Snow Day & integer in $[0, 20]$, float in $[0.0, 20.0]$, float in $[0.0, 10.0]$, float in $[0.0, 1.0]$ & float \\
    Solve Boolean & string of length $[1, 20]$ made of characters from \texttt{\{t, f, |, \&\}} & Boolean \\
    Spin Words & string of length $[0, 20]$ & string \\
    Square Digits & integer in $[0, 1000000]$ & string \\
    Substitution Cipher & 3 strings of length $[0, 26]$ & string \\
    Twitter & string of length $[0, 200]$ & string \\
    Vector Distance & 2 vectors of floats of length $[1, 20]$ with each float in $[-100.0, 100.0]$ & float \\
    \bottomrule
\end{tabular}
\end{table*}

\begin{enumerate}[itemsep=8pt,label=\arabic*.]
    \item \textbf{Basement (AoC)} Given a vector of integers, return the first index such that the sum of all integers from the start of the vector to that index (inclusive) is negative.
    \cite{prob:basement}
    
    \item \textbf{Bouncing Balls (CW)} Given a starting height and a height after the first bounce of a dropped ball, calculate the \textit{bounciness index} (height of first bounce / starting height). Then, given a number of bounces, use the bounciness index to calculate the total distance that the ball travels across those bounces.
    \cite{prob:bouncing-ball}

    \item \textbf{Bowling (CW)} Given a string representing the individual bowls in a 10-frame round of 10 pin bowling, return the score of that round.
    \cite{prob:bowling}

    \item \textbf{Camel Case (CW)} Take a string in \texttt{kebab-case} and convert all of the words to \texttt{camelCase}. Each group of words to convert is delimited by \texttt{"-"}, and each grouping is separated by a space. For example: \texttt{"camel-case example-test-string"} $\rightarrow$ \texttt{"camelCase exampleTestString"}.
    \cite{prob:camel-case}

    \item \textbf{Coin Sums (PE)} Given a number of cents, find the fewest number of US coins (pennies, nickles, dimes, quarters) needed to make that amount, and return the number of each type of coin as a separate output.
    \cite{prob:coin-sums}

    \item \textbf{Cut Vector (CW)} Given a vector of positive integers, find the spot where, if you cut the vector, the numbers on both sides are either equal, or the difference is as small as possible. Return the two resulting subvectors as two outputs.
    \cite{prob:cut-vector}

    \item \textbf{Dice Game (PE)} Peter has an $n$ sided die and Colin has an $m$ sided die. If they both roll their dice at the same time, return the probability that Peter rolls strictly higher than Colin.
    \cite{prob:dice-game}

    \item \textbf{Find Pair (AoC)} Given a vector of integers, return the two elements that sum to a target integer.
    \cite{prob:find-pair}

    \item \textbf{Fizz Buzz (CW)} Given an integer $x$, return \texttt{"Fizz"} if $x$ is divisible by 3, \texttt{"Buzz"} if $x$ is divisible by 5, \texttt{"FizzBuzz"} if $x$ is divisible by 3 and 5, and a string version of $x$ if none of the above hold.
    \cite{prob:fizz-buzz}

    \item \textbf{Fuel Cost (AoC)} Given a vector of positive integers, divide each by 3, round the result down to the nearest integer, and subtract 2. Return the sum of all of the new integers in the vector.
    \cite{prob:fuel-cost}

    \item \textbf{GCD [Greatest Common Divisor] (CW)} Given two integers, return the largest integer that divides each of the integers evenly.
    \cite{prob:gcd}

    \item \textbf{Indices of Substring (CW)} Given a text string and a target string, return a vector of integers of the indices at which the target appears in the text. If the target string overlaps itself in the text, all indices (including those overlapping) should be returned.
    \cite{prob:indices-of-substring}

    \item \textbf{Leaders (CW)} Given a vector of positive integers, return a vector of the leaders in that vector. A leader is defined as a number that is greater than or equal to all the numbers to the right of it. The rightmost element is always a leader.
    \cite{prob:leaders}

    \item \textbf{Luhn (CW)} Given a vector of 16 digits, implement Luhn's algorithm to verify a credit card number, such that it follows the following rules: double every other digit starting with the second digit. If any of the results are over 9, subtract 9 from them. Return the sum of all of the new digits.
    \cite{prob:luhn}

    \item \textbf{Mastermind (HW)} Based on the board game Mastermind. Given a Mastermind code and a guess, each of which are 4-character strings consisting of 6 possible characters, return the number of white pegs (correct color, wrong place) and black pegs (correct color, correct place) the codemaster should give as a clue.

    \item \textbf{Middle Character (CW)} Given a string, return the middle character as a string if it is odd length; return the two middle characters as a string if it is even length.
    \cite{prob:middle-character}

    \item \textbf{Paired Digits (AoC)} Given a string of digits, return the sum of the digits whose following digit is the same.
    \cite{prob:paired-digits}

    \item \textbf{Shopping List (CW)} Given a vector of floats representing the prices of various shopping goods and another vector of floats representing the percent discount of each of those goods, return the total price of the shopping trip after applying the discount to each item.
    \cite{prob:shopping-list}

    \item \textbf{Snow Day (HW)} Given an integer representing a number of hours and 3 floats representing how much snow is on the ground, the rate of snow fall, and the proportion of snow melting per hour, return the amount of snow on the ground after the amount of hours given. Each hour is considered a discrete event of adding snow and then melting, not a continuous process.

    \item \textbf{Solve Boolean (CW)} Given a string representing a Boolean expression consisting of \texttt{T}, \texttt{F}, \texttt{|}, and \texttt{\&}, evaluate it and return the resulting Boolean.
    \cite{prob:solve-boolean}

    \item \textbf{Spin Words (CW)} Given a string of one or more words (separated by spaces), reverse all of the words that are five or more letters long and return the resulting string.
    \cite{prob:spin-words}

    \item \textbf{Square Digits (CW)} Given a positive integer, square each digit and concatenate the squares into a returned string.
    \cite{prob:square-digits}

    \item \textbf{Substitution Cipher (CW)} This problem gives 3 strings. The first two represent a cipher, mapping each character in one string to the one at the same index in the other string. The program must apply this cipher to the third string and return the deciphered message.
    \cite{prob:substitution-cipher}

    \item \textbf{Twitter (HW)} Given a string representing a tweet, validate whether the tweet meets Twitter's original  character requirements. If the tweet has more than 140 characters, return the string \texttt{"Too many characters"}. If the tweet is empty, return the string  \texttt{"You didn't type anything"}. Otherwise, return \texttt{"Your tweet has X characters"}, where the $X$ is the number of characters in the tweet.

    \item \textbf{Vector Distance (CW)} Given two $n$-dimensional vectors of floats, return the Euclidean distance between the two vectors in $n$-dimensional space.
    \cite{prob:vector-distance}
    
\end{enumerate}

\section{Using PSB2}
\label{section:using-psb2}

While Section~\ref{section:problem-descriptions} provides English-language descriptions of the 25 benchmark problems, these are not sufficient to implement each problem in a new system. Here we discuss the system-agnostic details for using these problems.

For reasons discussed in Section~\ref{section:lessons}, we have created datasets consisting of large numbers of inputs and correct outputs for every problem~\cite{helmuth:psb2:zenodo}.\footnote{Our datasets follow the model of other machine learning datasets such as Penn ML Benchmarks~\cite{Olson2017PMLB, le2020pmlb} and the UCI ML Repository~\cite{Dua:2019}.} The dataset for each problem consists of a small number of hand-chosen inputs, often addressing edge cases for the problem, and 1 million randomly-generated inputs falling within the constraints of the problem. We recommend each different program synthesis run use a different set of data, composed of every one of the hand-chosen inputs and a random sample of the randomly-generated inputs. The alternative method of using the same fixed set of inputs for every run could happen to use a particularly lucky (or unlucky) set of inputs; using randomized inputs avoids this issue. Our datasets will allow those implementing PSB2 to simply sample the provided data, greatly decreasing the barrier to using PSB2. The PSB2 datasets can be found permanently on Zenodo.\footnote{\url{https://zenodo.org/record/4678739}} For more information about distributions of inputs in randomly-generated inputs, see the reference implementation, which was used to generate the datasets.\footnote{\url{https://github.com/thelmuth/Clojush/releases/tag/psb2-v1.0}}

When using our provided datasets, one could sample different sizes of training and unseen test sets to fit a given experiment. Our recommendation, which we use in the our experiments, is to use 200 example cases for the training set (including all hand-chosen inputs) and 2000 for the unseen test set. However, some synthesis methods may need smaller or larger training sets, and PSB2 can flexibly adapt to such systems. In order to produce fairer comparisons between systems, we recommend using a fixed program execution budget to limit the number of generated program executions in a single program synthesis run. We recommend a budget of 60 million program executions; we allocate these to 200 training cases used to evaluate a population of 1000 individuals for 300 generations in our experimental GP runs, but other allocations of the same executions would be reasonable.

Program synthesis methods that have been applied to PSB1 have used varying methods for constraining the instruction set and other program syntax. For example, some have used grammars~\cite{Forstenlechner:2017:EuroGP, Forstenlechner:2018:PPSN, Hemberg:2019:GECCO, Kelly:2019:EuroGP, ONeill:2019:CEC, Sobania:2020:EuroGP, Lynch:2020:PPSN} while others have used data-type categorized subsets of an instruction set~\cite{Helmuth:2015:GECCO, Helmuth:2020:ALife:source}. We do not want to constrain what a reasonable approach to selecting instructions may look like for any given program synthesis system. However, we also warn against cherry-picking a small subset of instructions suspected of being useful for a particular problem. Part of the difficulty of general program synthesis is that a system must manage a large set of potentially useful instructions, finding those relevant to a particular problem. We recommend employing a large set of general-purpose instructions when using PSB2 to benchmark program synthesis to best replicate the conditions of a real-world scenario.


When evaluating the performance of a synthesis system on PSB2, we recommend using \textit{success rate} (the number of synthesis runs that produce a solution) as the primary measure of performance, as was recommended in PSB1~\cite{Helmuth:2015:GECCO}.
For the synthesis of software, generating programs that pass most training cases is not sufficient; for this reason success rate is a better measure of performance than other metrics such as mean best fitness or mean number of training cases passed.
In particular, a solution must not only pass all cases in the training set, but also all of the cases in the test set, to ensure that it generalizes to unseen data. 
This avoids considering programs that overfit the training data, such as by memorizing the correct output to each input, as solutions.


\section{Experimental Methods and System Parameters}
\label{section:methods}


In this section we will discuss the system-specific parameters and choices that must be decided in order to use PSB2. In contrast with the previous section on general considerations, the choices here may differ considerably for different program synthesis systems. For our experiments, we used PushGP; we will describe in general the decisions that must be made and give our specific choices.

PushGP evolves programs in the language Push, a stack-based programming language built specifically for use in genetic programming~\cite{spector:2002:GPEM, 1068292}. Every data type has its own stack, and each Push instruction acts by pushing and popping various elements on and off the stacks. The output of each problem is typically the top element on a particular stack. The interpreter executes programs that are themselves placed on an \texttt{exec} stack, allowing \texttt{exec} instructions to manipulate control flow as well as the program itself as it runs. We provide a reference implementation in Clojure of the PushGP system used to produce our results, which includes each problem in PSB2.\footnote{\url{https://github.com/thelmuth/Clojush/releases/tag/psb2-v1.0}} This reference implementation is the same implementation of PushGP used in recent research using PSB1, e.g.~\cite{Helmuth:2020:ALife:downsampledlexicase, Helmuth:2018:GECCO, Helmuth:2017:GECCO}.

\begin{table*}
\centering
\caption{Instructions and data types used in our PushGP implementation of each problem. The column ``\# Instructions'' reports the number of instructions, terminals, and ephemeral random constants (ERC) used for each problem. The middle columns show which data types were used for each problem. For example, the Basement problem used all instructions relevant to exec, integers, Booleans, and vectors of integers. The last column lists the constants and ERCs used for the problem. Here, char constants are represented in the Clojure style, starting with a backslash, and strings are surrounded by double quotation marks. The ``Problems'' row simply counts how many problems use each data type. The ``Instructions'' row shows the number of Push instructions that primarily use each data type; some use multiple types but are only counted once.}
\label{table:DataTypes}
\rowcolors{2}{gray!15}{white}
\begin{tabular}{l r cccccccc p{8cm}}
    \toprule
    \textbf{Problem} & \begin{sideways}\textbf{\# Instructions} \end{sideways} & \begin{sideways}\textbf{exec}\end{sideways} & \begin{sideways}\textbf{integer}\end{sideways} & \begin{sideways}\textbf{float}\end{sideways} & \begin{sideways}\textbf{Boolean}\end{sideways} & \begin{sideways}\textbf{char}\end{sideways} &  \begin{sideways}\textbf{string}\end{sideways} & \begin{sideways}\textbf{vector of integers} \end{sideways} & \begin{sideways}\textbf{vector of floats}\end{sideways} &  \textbf{Constants and ERCs (besides inputs)} \tabularnewline
    \midrule
    Basement & 117 & x & x &  & x &  &  & x &  & \texttt{[], -1, 0, 1, integer ERC} \\
    Bouncing Balls & 127 & x & x & x & x &  &  &  &  & \texttt{0.0, 1.0, 2.0} \\
    Bowling & 161 & x & x &  & x & x & x &  &  & \texttt{\textbackslash -, \textbackslash X, \textbackslash /, \textbackslash 1, \textbackslash 2, \textbackslash 3, \textbackslash 4, \textbackslash 5, \textbackslash 6, \textbackslash 7, \textbackslash 8, \textbackslash 9, \textbackslash 10, integer ERC} \\
    Camel Case & 151 & x & x &  & x & x & x &  &  & \texttt{\textbackslash -, \textbackslash space, visible character ERC, string ERC} \\
    Coin Sums & 86 & x & x &  & x &  &  &  &  & \texttt{0, 1, 5, 10, 25} \\
    Cut Vector & 116 & x & x &  & x &  &  & x &  & \texttt{[], 0} \\
    Dice Game & 125 & x & x & x & x &  &  &  &  & \texttt{0.0, 1.0} \\
    Find Pair & 120 & x & x &  & x &  &  & x &  & \texttt{-1, 0, 1, 2, integer ERC} \\
    Fizz Buzz & 118 & x & x &  & x &  & x &  &  & \texttt{"Fizz", "Buzz", "FizzBuzz", 0, 3, 5} \\
    Fuel Cost & 117 & x & x &  & x &  &  & x &  & \texttt{0, 1, 2, 3, integer ERC} \\
    GCD & 79 & x & x &  & x &  &  &  &  & \texttt{integer ERC} \\
    Indices of Substring & 184 & x & x &  & x & x & x & x &  & \texttt{[], "", 0, 1} \\
    Leaders & 114 & x & x &  & x &  &  & x &  & \texttt{[], vector ERC} \\
    Luhn & 117 & x & x &  & x &  &  & x &  & \texttt{0, 2, 9, 10, integer ERC} \\
    Mastermind & 123 & x & x &  & x & x & x &  &  & \texttt{0, 1, \textbackslash B, \textbackslash R, \textbackslash W, \textbackslash Y, \textbackslash O, \textbackslash G} \\
    Middle Character & 151 & x & x &  & x & x & x &  &  & \texttt{"", 0, 1, 2, integer ERC} \\
    Paired Digits & 149 & x & x &  & x & x & x &  &  & \texttt{0, char digit ERC, integer ERC} \\
    Shopping List & 161 & x & x & x & x &  &  &  & x & \texttt{0.0, 100.0, float ERC} \\
    Snow Day & 131 & x & x & x & x &  &  &  &  & \texttt{0, 1, -1, 0.0, 1.0, -1.0} \\
    Solve Boolean & 153 & x & x &  & x & x & x &  &  & \texttt{true, false, \textbackslash t, \textbackslash f, \textbackslash \&, \textbackslash |} \\
    Spin Words & 152 & x & x &  & x & x & x &  &  & \texttt{4, 5, \textbackslash space, visible character ERC, string ERC} \\
    Square Digits & 151 & x & x &  & x & x & x &  &  & \texttt{"", 0, 1, 2, integer ERC} \\
    Substitution Cipher & 151 & x & x &  & x & x & x &  &  & \texttt{"", 0} \\
    Twitter & 153 & x & x &  & x & x & x &  &  & \texttt{0, 140, "Too many characters", "You didn't type anything", "your tweet has " , " characters"} \\
    Vector Distance & 160 & x & x & x & x &  &  &  & x & \texttt{[], 0} \\
    \midrule
    \rowcolor{white} Problems &  & 25 & 25 & 5 & 25 & 11 & 12 & 7 & 2 & \\
    \rowcolor{white} Instructions &  & 29 & 33 & 45 & 21 & 21 & 47 & 34 & 34 & \\
    \bottomrule
\end{tabular}
\end{table*}

We discuss the general design of program synthesis instruction sets in Section~\ref{section:using-psb2}. For our PushGP experiments, we use the general process recommended in PSB1, where, for each problem, we identify which data types (corresponding to stacks) are relevant and include all implemented instructions that use those stacks~\cite{Helmuth:2015:GECCO}. In Table~\ref{table:DataTypes}, we present the data types we chose to include for each problem, and the total number of instructions in the instruction set. These large instruction sets contain a wide range of general-purpose Push instructions, including some new instructions implemented since PSB1, avoiding the cherry-picking of clearly useful instructions. See the reference implementation for a complete listing of instructions.

Research utilizing PSB1 in using transfer-learned instruction sets showed that the composition of the instruction set matters a great deal to problem-solving performance~\cite{Helmuth:2020:ALife:source}. While we do not use fully transfer-learned instruction sets here, we do make use of one simple take-away: that including larger proportions of input instructions and constants/ERCs improves performance. An explanation of this result is that most Push instructions decrease stack sizes by consuming arguments and producing fewer return values, so increasing inputs and constants creates more data on which instructions can act. We boost the presence of input instructions and constants in the instruction set, making input instructions fill 15\% of the instruction set and constants/ERCs fill 5\% of the instruction set. The additional input instructions are evenly distributed between each input for problems with multiple inputs, and constants/ERCs are similarly evenly distributed for each listed in the last column of Table~\ref{table:DataTypes}.

For problems with multiple outputs, different synthesis systems will need to make choices specific to the language of the synthesized programs. Initial experiments in PushGP show that it achieves better results on multi-output problems when using one output instruction per output. These output instructions are included in the instruction set for such problems and will always appear in solution programs. An example of this is for Coin Sums, which has 4 outputs. We provide four corresponding output instructions, each of which takes the top integer from the integer stack and stores it in a write-only register for that output; further calls to an output instruction will overwrite this output register.

In order to define each problem for GP, we not only need the inputs and correct outputs for each problem, but also how to calculate the error function based on the correct output and a program's output. Here we describe the error functions we employed in our experiments, which we recommend for any GP system implementing PSB2; other non-GP program synthesis systems may require entirely different metrics. For each output data type, we use the following standard error functions for problems outputting that data type:
\begin{itemize}
    \item Integer or float: absolute value of the difference between program output and correct output.
    
    \item Boolean: 0 for correct and 1 for incorrect output.
    
    \item String: Levenshtein string edit distance between the program output and correct output.
    
    \item Vector of integers: add the difference in length between the program's output vector and the correct vector times 1000 to the absolute difference between each integer and the corresponding integer in the correct vector. 
\end{itemize}
The only exception is for the Indices of Substring problem, where we used Levenshtein distance to compare vectors of integers, since it makes more sense for that problem. In PushGP, some evolving programs will not return values of a program's output data type; we give a penalty error value specific to the problem when this occurs.

As has been shown to be effective at improving generalization, we use an automatic simplification procedure on every evolved Push program that passes all of the training cases before testing it on the test set~\cite{Helmuth:2017:GECCO}.

Unlike for PSB1, we aimed to keep all system-specific parameters constant between problems, increasing ease of use for both implementation and reporting of results. These parameters were chosen based on prior experience and reasonable performance; we leave optimizing parameter settings as an open research question. Other systems may choose to use different system-specific parameters.

Our PushGP system uses linear Plush genomes that are initialized by generating lists of random instructions from the instruction set~\cite{Helmuth:2016:GPTP}. We list the important parameters used in our experiments below:
\begin{itemize}
\item 
Maximum initial genome size: 250 genes
\item 
Maximum genome size: 500 genes
\item 
Population size: 1000
\item 
Maximum generations per run: 300
\item 
Maximum steps of the Push interpreter when executing one program: 2000
\item 
Parent selection: lexicase selection~\cite{Helmuth:2015:ieeeTEC}
\item 
Genetic operator: Uniform Mutation with Additions and Deletions (UMAD), used to make 100\% of children~\cite{Helmuth:2018:GECCO}.
\item 
UMAD addition rate: 0.09
\end{itemize}
As described in Section~\ref{section:using-psb2}, using the exact same population size and generations is not necessary for comparisons between systems; instead, we recommend using a maximum budget of 60 million program executions regardless of other settings.

\section{Experimental Results}
\label{sec:results}

\begin{table}[t]
\centering
\caption{Results from 100 PushGP runs on each problem. ``Succ.'' gives the number of runs that successfully find a program that pass every training case and perfectly pass a set of 2000 unseen test cases. ``Gen.'' gives the proportion of solutions on the training data that generalize to unseen data. ``Size'' gives the size of the smallest automatically simplified solution that generalized to unseen data. Time is the average number of seconds taken per generation.}
\label{table:results}
\rowcolors{2}{gray!15}{white}
\begin{tabular}{lrrrr}
    \toprule
    \textbf{Problem} & \textbf{Succ.} & 
    \textbf{Gen.} & \textbf{Size} & \textbf{Time} \\
    \midrule
    Basement & 1 & 1.00 & 18 & 250 \\
    Bouncing Balls & 0 & 0.00 & - & 311 \\
    Bowling & 0 & - & - & 206 \\
    Camel Case & 1 & 1.00 & 20 & 95\\
    Coin Sums & 2 & 1.00 & 33 & 213 \\
    Cut Vector & 0 & - & - & 194 \\
    Dice Game & 0 & - & - & 287 \\
    Find Pair & 4 & 1.00 & 16 & 763 \\
    Fizz Buzz & 25 & 0.96 & 19 & 281 \\
    Fuel Cost & 50 & 1.00 & 9 & 305 \\
    GCD & 8 & 0.67 & 19 & 198 \\
    Indices of Substring & 0 & - & - & 241 \\
    Leaders & 0 & - & - & 302 \\
    Luhn & 0 & - & - & 239 \\
    Mastermind & 0 & - & - & 126 \\
    Middle Character & 57 & 0.86 & 10 & 547 \\
    Paired Digits & 8 & 1.00 & 15 & 250 \\
    Shopping List & 0 & - & - & 714 \\
    Snow Day & 4 & 1.00 & 11 & 263 \\
    Solve Boolean & 5 & 1.00 & 18 & 373 \\
    Spin Words & 0 & - & - & 443 \\
    Square Digits & 0 & - & - & 435 \\
    Substitution Cipher & 60 & 0.98 & 9 & 395 \\
    Twitter & 31 & 0.74 & 22 & 527 \\
    Vector Distance & 0 & - & - & 667 \\
    \bottomrule
\end{tabular}
\end{table}

In order to give a baseline performance of the 25 problems in PSB2, we conducted 100 PushGP runs on each problem using the experimental methods described in Section~\ref{section:methods}. We present success rates of our runs in Table~\ref{table:results}. Out of the 25 problems in PSB2, 13 were solved by PushGP. Of these 13, 3 of them had 50 or more successes (Fuel Cost, Middle Character, and Substitution Cipher) and 2 others had 25 or more successes (Fizz Buzz and Twitter). The remaining 8 had fewer than 10 solutions, showing that they are solvable by GP but leave a lot of room for improvement.

PushGP did not solve the remaining 12 problems. However, we note that in our initial exploratory runs of the Bouncing Balls and Leaders problems, PushGP produced 2 generalizing solutions to each, but did not replicate these successes in our runs with finalized parameter settings. Additionally, in continued work using the same PushGP settings as this paper except using down-sampled lexicase selection~\cite{Helmuth:2020:ALife:downsampledlexicase}, small numbers of generalizing solutions were found to the problems Bouncing Balls, Dice Game, Indices of Substring, and Square Digits problems~\cite{Helmuth:2021:ALifeJournal}.
Thus at least 18 of the 25 problems are solvable with the our PushGP implementation. While we have no guarantees that the other 7 problems can be solved by any program synthesis system, they provide useful targets for future research.

The second column in Table~\ref{table:results} gives the generalization rate of all evolved solutions for problems on which PushGP produced at least one program that solved every training case. The generalization rate is calculated as the number of solution programs that pass the unseen test set divided by the number of solution programs that pass the training set. For most problems with training set solutions, those solutions tended to generalize well with rates of 0.95 to 1.0. The three problems with lower generalization rates, GCD, Middle Character, and Twitter all had rates over 0.5. However, Bouncing Balls found 2 solutions on the training data, but neither of them generalized to the test, which resulted in 0 successful runs and a 0.00 generalization rate.

Another way of approximating the difficulty of these problems is by looking at the size of the smallest solution program found for each problem. Smaller solutions are easier for a program synthesis system to generate, simply because they require assembling fewer instructions in the right order. Our results are particular to Push program solutions, but should correlate with the sizes of programs needed to solve these problems in other systems. In order to find each size, we took each solution program and automatically simplified it to produce a smaller equivalent program~\cite{Helmuth:2017:GECCO}. Of these simplified programs, in Table~\ref{table:results} we report the smallest simplified solution size out of all simplified solutions to each problem. We see that the smallest solution size is 9 instructions for two problems, and two others have sizes of 10 and 11; three of these problems also had the highest success rates in PSB2. Many others have larger smallest solution sizes, though we note that with the small sample sizes of solutions for some problems, smaller solutions may exist. In comparison, \cite{Helmuth:2015:GECCO} reported that 8 of the problems in PSB1 had a smallest solution size less than 9, the minimum for PSB2. Along with success rates, these sizes of smallest solutions give evidence that the problems in PSB2 are more difficult than those in PSB1.

The last column of Table~\ref{table:results} gives the average number of seconds per generation over all of the PushGP runs for the problem. Note that these runs were conducted on two different computing clusters, each of which is composed of heterogeneous machines, so these measurements should only be considered as rough approximations of running time. To that end, we note that all problems have generational running times within one order of magnitude of each other, meaning there are not any exceptionally slow or fast problems.

\section{Conclusions}

We have presented PSB2, the second generation of general program synthesis benchmark problems. We discussed the past research that has used PSB1, the lessons learned from years of its use, and why we need a new benchmark suite. We then provided the sources and problems that make up PSB2, giving details of how to implement and use it in new systems. Finally, we presented experimental results showing the increased difficulty of the problems of PSB2 compared to PSB1.

After the results we have presented using PushGP, we anticipate using other GP systems (such as those we mention in Section~\ref{section:past-research} that have used PSB1) to produce initial results on PSB2 will provide a useful comparison. Additionally, we encourage the application of non-evolutionary automatic program synthesis methods to these problems, to better gauge the strengths and weaknesses of these different methods.

The lessons learned from PSB1 will make it easier to implement PSB2 in new program synthesis systems, increasing adoption in the community and furthering the field. PSB2 will provide a new target for program synthesis systems, stretching their capabilities and moving the field toward the types of problems that may be encountered in real-world program synthesis applications.

\begin{acks}
  The authors would like to thank Lee Spector, Grace Woolson, and Amr Abdelhady for discussions that helped shape this work.
\end{acks}

\bibliographystyle{ACM-Reference-Format}
\bibliography{bib-benchmarks}

\end{document}